\documentclass{ctr_summer}

\usepackage{ctrfont}

\usepackage[dvips]{graphicx}
\usepackage{psfrag}
\usepackage{amsmath,rotating}
\usepackage{natbib}

\title{Wave propagation in the magnetic sun}
\shorttitle{Wave propagation in the magnetic sun}

\author{T.~Hartlep, M.~S.~Miesch\footnote{High Altitude Observatory, National Center for Atmospheric Research, Boulder, CO}, \and N.~N.~Mansour}

\shortauthor{ T.~Hartlep, M.~S.~Miesch, \& N.~N.~Mansour}

\begin{document}

\maketitle

This paper reports on efforts to simulate wave propagation in the solar interior. 
Presented is work on extending a numerical code for constant entropy acoustic waves in the absence of magnetic fields to the case where magnetic fields are present. 
A set of linearized magnetohydrodynamic (MHD) perturbation equations has been derived and implemented.

\vskip0.1in
\hrule

\section{Introduction}
\label{Section:Introduction}

The evolution of the solar interior is
actively studied
theoretically, numerically, and observationally.
Many recent advancements in our understanding of the structure and dynamics of the sun have been made by studying solar oscillation.
This is the science of helioseismology.
Through ground- and satellite-based telescopes, oscillations on the solar surface are being observed and, through appropriate techniques, are used to infer information about the sun's internal structure and composition.
One interesting technique provides so-called far-side images of the sun \cite*[]{Lindsey00},
where maps of active regions on the far side (the side of the sun facing away from earth) are inferred from the oscillations observed on the front side (the one facing earth).
For a review of other helioseismic methods, as well as general properties of the observed solar oscillations, 
see \cite{ChristensenDalsgaard02}.
In general, these inferences are based on simplified models of wave propagation, and researchers in the field \cite*[e.g.,][]{Werne04} have expressed their need for wave propagation simulations to test and calibrate these methods.

Of course, wave propagation can be studied by simulating the full compressible
equations; for instance, simulations of the shallow upper layer of the solar convection zone by \cite{Stein00} demonstrate excellent agreement with existing analytical theories and observations.
But due to the enormous computational requirements, these types of simulations are able to simulate only a very small part of the sun.
For global simulations such an approach is not expected to be feasible in the immediate future.
Except for the very near surface region, though, flow velocities in the sun are much smaller than the speed of sound and a perturbation approach is applicable.
Thus, acoustic waves can be treated here as small perturbations traveling through a base flow.
This is the foundational idea of the this project.
The base flow itself can be either artificially prescribed or it can be provided by other simulations such as three-dimensional global simulations of solar convection in the anelastic approximation  \cite*[]{Miesch98,Brun02,Brun04}.
Our work extends that of~\cite{Hartlep05_1} by including the effects of a prescribed magnetic field on the propagation of waves, which is especially of interest for testing the far-side imaging technique previously mentioned.
We present here the perturbation equations we derived and our numerical method, and comment on the current status of this project.





\section{Model equations}

The idea underlying the derivation is to split the dependent variables (density, pressure, magnetic field strength, etc.) into base variables, denoted by $\tilde{\cdot}$, and wave perturbations, $\cdot^\prime$, and then to derive appropriate equations for the perturbations. 
We will take the base state to be non-rotating, without flows ($\boldsymbol{\tilde{v}}=0$), and with prescribed three-dimensional sound speed $c_s$, density $\tilde{\rho}$, magnetic field $\boldsymbol{\tilde{B}}$, and electric current distribution $\boldsymbol{\tilde{J}}$.
The real sun, of course, is rotating and has non-vanishing flow velocities.
We plan to add these influences at a later stage,
but the current focus is 
on the effects that magnetic fields and spatial variations in the speed of sound have on the propagation of acoustic waves. 
The linearized perturbation equations arising from the continuity and Euler's equation are:
\begin{eqnarray}
 \frac{\partial}{\partial t} \rho^\prime & = & - \boldsymbol{\nabla} \cdot \boldsymbol{m}^\prime
 \label{Eqn:rhoprime} \\
 \frac{\partial}{\partial t} \boldsymbol{m}^\prime & = & - \boldsymbol{\nabla} p^\prime + \rho^\prime  \boldsymbol{\tilde{g}} + \boldsymbol{L}^\prime
  \label{Eqn:mprime}
\end{eqnarray}
with $\boldsymbol{m}^\prime=\tilde{\rho} \boldsymbol{v}^\prime$,  $\boldsymbol{\tilde{g}}=-\boldsymbol{\hat{r}}\tilde{g}$, and $\boldsymbol{L}^\prime$ being the mass flux, the acceleration due to gravity, and the perturbation of the Lorentz force, respectively.
The background state is assumed to be in magnetohydrostatic balance: $\boldsymbol{\nabla}\tilde{p}=-\tilde{\rho}\boldsymbol{\tilde{g}}+\boldsymbol{\tilde{L}}$. 
In principle, another term in the momentum equation~(\ref{Eqn:mprime}) appears to account for the perturbation of the gravitational acceleration caused by the waves, $-\tilde{\rho}\boldsymbol{g}^\prime$.
This term can be expected to be very small, though, and is neglected here.
Needed next is an equation that relates the pressure to the density perturbation.
Assuming adiabatic processes, we have $p^\prime = c_s^2 \rho^\prime + \gamma \tilde{p} / C_p s^\prime$ with $C_p$ and $\gamma$ being the specific heat at constant pressure and the adiabatic index, respectively.
Only isentropic waves are considered at this point, and therefore $s^\prime$, the entropy perturbation, is set to zero.
The magnetic field, of course, is divergence free,
\begin{equation}
 \boldsymbol{\nabla} \cdot \boldsymbol{B}^\prime = 0,
\end{equation}
and is given by Faraday's law of induction:
\begin{equation}
\frac{\partial}{\partial t} \boldsymbol{B}^\prime = - c \boldsymbol{\nabla}\times\boldsymbol{E}^\prime.
\end{equation}
Also included are Ohm's law (in the infinite conductivity limit), Ampere's law, and the Lorentz force given by:
\begin{eqnarray}
  \boldsymbol{E}^\prime & = & - \frac{1}{c} \boldsymbol{v}^\prime \times \boldsymbol{\tilde{B}},
  \label{Eqn:ohmslaw} \\
  \boldsymbol{J}^\prime & = & \frac{c}{4\pi} \boldsymbol{\nabla}\times\boldsymbol{B}^\prime,
  \label{Eqn:ampereslaw} \\
  \boldsymbol{L}^\prime & = & \frac{1}{c} \big\{ \boldsymbol{J}^\prime \times \boldsymbol{\tilde{B}} + \boldsymbol{\tilde{J}} \times \boldsymbol{B}^\prime \big\}.
 \label{Eqn:lorentzforce}
\end{eqnarray}
The displacement current $\partial_t \boldsymbol{E}^\prime$ in the equation for the electric current has been neglected because the velocities involved are small compared to the speed of light $c$.
These linearized equations do not account for the initial excitation of the waves, which is a non-linear process.
Acoustic waves are expected to be generated by the vigorous turbulent convection close below the solar surface.
We simulate these stochastic excitations by artificially adding a random forcing term to Eq.~(\ref{Eqn:rhoprime}).

\section{Numerical method}

The preceding equations are solved in spherical coordinates with a pseudo-spectral method.
Scalar quantities such as pressure and density are expanded in terms of spherical harmonic basis functions for their angular structure, and B-splines~\cite*[]{DeBoor87,Loulou97,Hartlep04_3,Hartlep05_1} for their radial dependence.
Vector fields such as the magnetic field, the Lorentz force, and the mass flux are expanded in vector spherical harmonics and B-splines. 
Vector spherical harmonics were selected because of the coordinate singularities in spherical coordinates, which are most easily treated in those basis functions (see Appendix~\ref{Section:Appendix}). 
Typically, B-splines of polynomial order 4 are used, and the spacing of the generating knot points is chosen to be proportional to the local speed of sound. 
This results in higher radial resolution near the solar surface, where the sound speed is low (less than 7~km/s) compared to the deep interior, where the sound speed surpasses 500~km/s.
The numerical domain is chosen to be larger than the solar radius of approximately 696~Mm by an additional 2--4~Mm to account for an absorption layer as a means of realizing non-reflecting boundary conditions.
This layer is implemented by adding damping terms $- \sigma \rho^\prime$ and $- \sigma \boldsymbol{m}^\prime$ to Eq.~(\ref{Eqn:rhoprime}) and~(\ref{Eqn:mprime}), where $\sigma$ is a positive damping coefficient that is independent of time, which is set to be zero in the interior, and from the solar surface up increases smoothly into the buffer layer.

The equations are then recast using an integrating factor $\exp(\sigma t)$, and advanced using a staggered leapfrog scheme in which $\rho^\prime $ and $\boldsymbol{B}^\prime$  are advanced at the same time, and $\boldsymbol{m^\prime}$ is offset by half of a time step.
The basic properties of the background state, $\tilde{\rho}$, $\tilde{g}$, and $c_s$, are obtained from solar models, which provide shell-averaged properties of the quiet sun.
Solar model S by \cite{ChristensenDalsgaard96} is used for radii below 696.5~Mm, the highest radius considered in that model.
Properties above this radius are taken from chromosphere model C by \cite{Vernazza81}.
Models for 3-D spatial variations of the temperature, and therefore the sound speed, and the magnetic field, in structures such as sunspots are then added to this basic, spherically symmetric background.

%
%

\section{Summary}
\label{Section:Future}

We have derived a set of linearized MHD perturbation equations for the propagation of waves in a simplified sun that is non-rotating and has no background flows, and we have implemented these equations as an extension to our previous non-magnetic wave propagation code~\cite[]{Hartlep05_1}.
This represents a considerable extension and required developing routines for vector spherical harmonics, including fast transformations between spectral and physical space (which were not present in the original code).
The new implementation is still in the testing phase, 
but we hope to be able to perform numerical experiments (e.g., a test of the far-side imaging technique) soon. 

\appendix

\section{Treatment of coordinate singularities in spherical coordinates}
\label{Section:Appendix}

The sun is very well approximated by a sphere, and spherical coordinates are therefore the obvious choice in the numerical method.
This requires, however, that special care is taken to ensure regularity of physical quantities on the polar axis and at the coordinate center.
The use of a spherical harmonics expansion guarantees smoothness at the polar axis, but additional constraints on the expansion arise from requiring regularity at the origin.
Fortunately, in many applications the origin can be excluded from the numerical domain and therefore those constraints are irrelevant.  
In the present case though, acoustic waves travel through the whole sphere, and the center cannot be excluded, requiring that we enforce these constraints.
The constraints are different for scalars such as density and pressure, and for vector quantities such as velocity and magnetic field. The following briefly outlines their derivation.

We start with a scalar variable, which is expanded in terms of spherical harmonics $Y_{l,m}$ for its angular dependence and, for each spherical degree $l$ and azimuthal index $m$, some radial functions $f_{l,m}$:
\begin{eqnarray}
  \label{Eqn:Singularity:ScalarExpansion}
   f(r,\theta,\phi) & = & \sum_{l=0}^{\infty} \sum_{m=-l}^{+l} f_{l,m}(r) \, Y_{l,m}(\theta,\phi).
\end{eqnarray}
Of course, in the actual implementation the infinite sum is truncated at a chosen maximum value of l.
Using the definition of the spherical harmonics and the transformation to Cartesian coordinates ($x = r \sin\theta \cos\phi$, $y = r \sin\theta \sin\phi$, $z = r \cos\theta$) we can rewrite Eq.~(\ref{Eqn:Singularity:ScalarExpansion}) into:
\begin{equation}
   \label{Eqn:Singularity:ScalarExpansion2}
   f(r,\theta,\phi) = \sum_{l=0}^{\infty} \sum_{m=-l}^{+l} \sum_{k=k_{min}}^l c_{k,l,m} (x \pm i y)^{|m|}  z^{2k-(l+|m|)} r^{-2k+l} f_{l,m}(r),
\end{equation}
with some constants $c_{k,l,m}$, and where $k_{min}=(l+|m|)/2$ for even values of $l+|m|$, and $k_{min}=(l+|m|+1)/2$ for odd values of $l+|m|$.
The plus and minus signs in Eq.~(\ref{Eqn:Singularity:ScalarExpansion2}) are used for positive and negative values of $m$, respectively.
The terms involving $x$, $y$, and $z$ are always regular at the center since $|m|$ and $2k-(l+|m|)$ are non-negative.
The only constraint is that $r^{-2k+l} f_{l,m}(r)$ must also be regular.
This function, expanded in a Taylor series around $r=0$, reads as:
\begin{equation}
  r^{-2k+l} f_{l,m}(r) = \sum_{p=0}^{\infty} \alpha_{l,m,p} r^{-2k+l} r^p,
\end{equation}
where individual terms are only smooth at the origin if the combined exponent $-2k+l+p$ is positive and even.
For all terms for which this is not the case, the expansion coefficient $\alpha_{l,m,p}$ must vanish. 
In other words, the constraint is that the radial function $f_{l,m}$ must be of the form:
\begin{equation}
  \label{Equation:Regularity:Scalar:Condition}
  f_{l,m}(r) = r^l P(r^2),
\end{equation}
with $P(r^2)$ being a smooth function in $r^2$.
An alternative derivation of this constraint can be found in~\cite{Stanaway88}.
For similar constraints that arise in Fourier expansions in cylindrical coordinates see \cite{Lewis90}.

Additional complications arise for vector quantities from the choice of unit vectors.
A rather simplistic way would be to write a vector field in terms of $r$-, $\theta$-, and $\phi$-components, and then to expand these components like scalars individually in spherical harmonic functions, i.e.,
\begin{equation}
  \label{Equation:Regularity:rthetaphi}
   \boldsymbol{F}(r,\theta,\phi) = \sum_{l=0}^{\infty} \sum_{m=-l}^{+l} Y_{l,m}(\theta,\phi) 
      \big\{ f_{\boldsymbol{r},l,m}(r) \boldsymbol{\hat{r}}(\theta,\phi) + f_{\boldsymbol{\theta},l,m}(r) \boldsymbol{\hat{\theta}}(\theta,\phi) + f_{\boldsymbol{\phi},l,m}(r) \boldsymbol{\hat{\phi}}(\theta,\phi) \big\}.
\end{equation}
Such an expansion is fine if the computational domain does not include the origin $r=0$, as is the case in the simulations of the solar convection layer by~\cite{Brun04}.
In our case, however, it is not an appropriate choice, since, for the vector field to be regular at the origin, $f_{r,l,m}$, $f_{\theta,l,m}$, and $f_{\phi,l,m}$ are not independent of each other and the regularity condition becomes hard to enforce.
The alternative here is to use so-called vector spherical harmonics as angular basis function:
\begin{equation}
   \boldsymbol{F}(r,\theta,\phi) = \sum_{l=0}^{\infty} \sum_{m=-l}^{+l}
      \big\{  
           f_{\boldsymbol{X},l,m}(r) \boldsymbol{X}_{l,m}(\theta,\phi) +
           f_{\boldsymbol{V},l,m}(r) \boldsymbol{V}_{l,m}(\theta,\phi) +
           f_{\boldsymbol{W},l,m}(r) \boldsymbol{W}_{l,m}(\theta,\phi)
       \big\},
\end{equation}
where $\boldsymbol{X}_{l,m}$, $\boldsymbol{V}_{l,m}$ and $\boldsymbol{W}_{l,m}$ are defined according to \cite{Hill54} by
\begin{eqnarray}
  \boldsymbol{X}_{l,m}  (\theta,\phi)& = & - \boldsymbol{\hat{\theta}}  \sqrt{\frac{1}{l(l+1)}} \frac{m Y_{l,m} (\theta,\phi)}{\sin\theta} 
		         -   \boldsymbol{\hat{\phi}} \sqrt{\frac{1}{l(l+1)}} i \frac{\partial}{\partial \theta} Y_{l,m} (\theta,\phi),
 \label{Eqn:Hill:X} \\
  \boldsymbol{V}_{l,m} (\theta,\phi)& = & - \boldsymbol{\hat{r}}  \sqrt{ \frac{l+1}{2l+1} } Y_{l,m} (\theta,\phi) 
                         +   \boldsymbol{\hat{\theta}}  \sqrt{\frac{1}{(l+1)(2l+1)}}  \frac{\partial}{\partial \theta} Y_{l,m} (\theta,\phi)   \nonumber \\
		       & + & \boldsymbol{\hat{\phi}}  \sqrt{\frac{1}{(l+1)(2l+1)}} \frac{i m Y_{l,m} (\theta,\phi)}{\sin\theta}, 
  \label{Eqn:Hill:V} \\
  \boldsymbol{W}_{l,m}  (\theta,\phi)& = & \boldsymbol{\hat{r}}  \sqrt{ \frac{l}{2l+1} } Y_{l,m} (\theta,\phi)  
                         +   \boldsymbol{\hat{\theta}}  \sqrt{\frac{1}{l(2l+1)}}  \frac{\partial}{\partial \theta} Y_{l,m} (\theta,\phi)  \nonumber \\
		       & + & \boldsymbol{\hat{\phi}} \sqrt{\frac{1}{l(2l+1)}} \frac{i m Y_{l,m} (\theta,\phi)}{\sin\theta}.
  \label{Eqn:Hill:W}
\end{eqnarray}
%
Here,  the regularity conditions for the radial functions decouple and are actually very similar to what is found for scalar fields. 
For instance, $\boldsymbol{X}_{l,m}$ can be rewritten as
\begin{eqnarray}
  \boldsymbol{X}_{l,m}(\theta,\phi) & = &
       (\boldsymbol{\hat{x}} + i \boldsymbol{\hat{y}}) \frac{1}{2} \sqrt{ (l+m)(l-m+1) } Y_{l,m-1}(\theta,\phi)    \nonumber \\
      & + & (\boldsymbol{\hat{x}} - i \boldsymbol{\hat{y}}) \frac{1}{2} \sqrt{ (l-m)(l+m+1) } Y_{l,m+1}(\theta,\phi)   \nonumber \\
     & + & \boldsymbol{\hat{z}} m Y_{l,m}(\theta,\phi),
\end{eqnarray}
and therefore, following the arguments above, the radial function $f_{\boldsymbol{X},l,m}$ must behave just like a scalar, i.e.,
\begin{equation}
  \label{Equation:Regularity:VSH:ConditionX}
  f_{\boldsymbol{X},l,m}(r) = r^l P_{\boldsymbol{X},l,m}(r^2),
\end{equation}
with again $P_{\boldsymbol{X},l,m}(r^2)$ being a smooth function in $r^2$.
The results for the other two functions are:
\begin{eqnarray}
  \label{Equation:Regularity:VSH:ConditionV}
  f_{\boldsymbol{V},l,m}(r)  & = & r^{l+1} P_{\boldsymbol{V},l,m}(r^2),  \\ 
  \label{Equation:Regularity:VSH:ConditionW}
  f_{\boldsymbol{W},l,m}(r)  & = & r^{l-1} P_{\boldsymbol{W},l,m}(r^2).
\end{eqnarray}
In our numerical method, all radial functions are expanded in B-splines, which are piecewise polynomials.
Conditions~(\ref{Equation:Regularity:Scalar:Condition}) and~(\ref{Equation:Regularity:VSH:ConditionX})--(\ref{Equation:Regularity:VSH:ConditionW}) result in a linear coupling of the expansion coefficients for B-splines near the center.
These linear systems are very small and easy to implement.
Only the first $k+1$ coefficients are coupled, where $k$ is the polynomial order of the B-splines used (here usually $k=4$), since these are the only B-splines that are non-zero close to boundary $r=0$.


\vspace{1cm}

\bibliography{hartlep}
\bibliographystyle{jfm}


\end{document}